# Concussion classification via deep learning using whole-brain white matter fiber strains

Yunliang Cai [1], Shaoju Wu [1], Wei Zhao [1], Zhigang Li [2], Songbai Ji [1, 3, 4*]

[1] Department of Biomedical Engineering, Worcester Polytechnic Institute, Worcester, MA 01605, USA

[2] Department of Biomedical Data Science, Geisel School of medicine, Dartmouth College, Hanover, NH 03755, USA

[3] Department of Mechanical Engineering, Worcester Polytechnic Institute, Worcester, MA 01609, USA

[4] Thayer School of Engineering, Dartmouth College, Hanover, NH 03755, USA

[*] Corresponding author: sji@wpi.edu

**Abstract**

Developing an accurate and reliable injury predictor is central to the biomechanical studies of traumatic brain injury. State-of-the-art efforts continue to rely on empirical, scalar metrics based on kinematics or model-estimated tissue responses explicitly pre-defined in a specific brain region of interest. They could suffer from loss of information. A single training dataset has also been used to evaluate performance but without cross-validation. In this study, we developed a deep learning approach for concussion classification using implicit features of the entire voxel-wise white matter fiber strains. Using reconstructed American National Football League (NFL) injury cases, leave-one-out cross-validation was employed to objectively compare injury prediction performances against two baseline machine learning classifiers (support vector machine (SVM) and random forest (RF)) and four scalar metrics via univariate logistic regression (Brain Injury Criterion (BrIC), cumulative strain damage measure of the whole brain (CSDM-WB) and the corpus callosum (CSDM-CC), and peak fiber strain in the CC). Feature-based deep learning and machine learning classifiers consistently outperformed all scalar injury metrics across all performance categories in cross-validation (e.g., average accuracy of 0.844 vs. 0.746, and average area under the receiver operating curve (AUC) of 0.873 vs. 0.769, respectively, based on the testing dataset). Nevertheless, deep learning achieved the best cross-validation accuracy, sensitivity, and AUC (e.g., accuracy of 0.862 vs. 0.828 and 0.842 for SVM and RF, respectively). These findings demonstrate the superior performances of deep learning in concussion prediction, and suggest its promise for future applications in biomechanical investigations of traumatic brain injury.

**Keywords:** Concussion classification, deep learning, fiber strain, white matter, Worcester Head Injury Model

**Introduction**

Traumatic brain injury (TBI) resulting from blunt head impact is a leading cause of morbidity and mortality in the United States [1]. The recent heightened public awareness of TBI, especially of sports-related concussion [2,3], has prompted the Institute of Medicine and National Research Council of the National Academies to recommend immediate attention to address the biomechanical determinants of injury risk and to identify effective concussion diagnostic metrics and biomarkers, among others [4].

Impact kinematics such as linear and rotational accelerations are convenient ways to characterize impact severity. Naturally, these simple kinematic variables and their more sophisticated variants have been used to assess the risk and severity of brain injury. As head rotation is thought to be the primary mechanism for mild TBI (mTBI) including sports-related concussion, most kinematics metrics include rotational acceleration or velocity, either solely (e.g., rotational injury criterion (RIC), power rotational head injury criterion (PRHIC) [5], brain injury criterion (BrIC) [6], and rotational velocity change index (RVCI) [7]) or in combination with liner acceleration [8].

Kinematic variables, alone, do not provide regional brain mechanical responses thought to cause injury [9]. Validated computational models of the human head are, in general, believed to serve as an important bridge between external impact and tissue mechanical responses. Model-estimated, response-based injury metrics are desirable, as they can be directly related to tissue injury tolerances. Commonly used tissue response metrics include peak maximum principal strain and cumulative strain damage measure (CSDM; [10]) for the whole brain. More recently, white matter (WM) fiber strain [11–14] is also being explored as a potential improvement. There is growing interest in utilizing model-simulated responses to benchmark the performance of other kinematic injury metrics [6,15–17].

Regardless of these injury prediction approaches (kinematic or response-based), they share some important common characteristics. First, they have utilized a single injury dataset for "training" and performance evaluation. Often, this was performed by fitting a univariate logistic regression model to report the area (AUC) under the receiver operating curve (ROC) [8,12,13,18]. However, without cross-validation using a separate "testing dataset", there could be uncertainty how the metrics perform when they are, presumably, deployed to predict injury on fresh, unmet impact cases [12,19]. This is an important issue seemingly under-appreciated, given that AUC especially from a single training dataset provides an average or aggregated performance of a procedure but does not directly govern how a clinical decision, in this case, injury vs. non-injury diagnosis, is made.

Second, an explicit, pre-defined kinematic or response metric is necessary for injury prediction. While candidate injury metrics are typically from known or hypothesized injury mechanisms (e.g., strain), they are derived empirically. For response-based injury metrics, they are also pre-defined in a specific brain region of interest (ROI) such as the corpus callosum (CC) and brainstem. However, they do not consider other anatomical regions or functionally important neural pathways. The commonly used peak maximum principal strain and CSDM describe the peak response in a single element or the volume fraction of regions above a given strain threshold, respectively. However, they do not (and cannot) inform the location or distribution of brain strains that are likely critical for concussion, given the widespread neuroimaging alterations [20] and a diverse spectrum of clinical signs and symptoms [21] observed in the clinic.

Consequently, even when using the same reconstructed American National Football League (NFL) head impacts, studies have found inconsistent "optimal" injury predictors (e.g., maximum shear stress in the brainstem [22], strain in the gray matter and $CSDM_{0.1}$ (using a strain threshold of 0.1) in the WM [18], peak axonal strain within the brainstem [13], or tract-wise injury susceptibilities in the super longitudinal fasciculus [14]). Most of these efforts are essentially "trial-and-error" in nature as they attempt to pinpoint a specific variable in a given ROI for injury prediction. However, no consensus has reached on the most injury discriminative metric or ROI. Without accounting for the location and distribution of brain responses that are likely critical to concussion, critical information is lost.

Injury prediction is a binary classification. Besides univariate logistic regression, there have been numerous algorithmic advances in classification, including feature-based machine learning and, more recently, deep learning [23,24]. Instead of relying on a single, explicit scalar metric that could suffer from loss of information, feature-based machine/deep learning techniques employs multiple features to perform classification. However, despite their successes [23,24], application of feature-based machine/deep learning in TBI biomechanics for injury diagnosis is extremely limited or even non-existent at present. A recent study utilized SVM to predict concussion [25]. However, it was limited to kinematic variables (vs. brain responses) and two injury cases, which did not allow for cross-validation.

Conventional machine learning classifiers such as support vector machine (SVM) and random forest (RF) have been widely used in medical imaging [26,27] and computer vision [28] applications. Deep learning is the most recent advancement in feature-based classification, and it has achieved remarkable success in a wide array of science domains (see [23] for a recent review). This technique has already been successfully applied in numerous neuroimaging analyses, including registration [29], segmentation [30], and WM fiber clustering based on learned shape features [31]. For neurological disease diagnosis, applications include the use of Convolutional Neural Network (CNN) for Alzheimer's detection [32], and fully connected Restricted Boltzmann Machine to detect mTBI categories based on diffusion tensor imaging (DTI) parameters [33]. However, this technique has not been employed for TBI prediction using brain tissue mechanical responses such as WM fiber strain (i.e., stretch along WM fiber directions). Unlike conventional neuroimages where tissue boundaries readily serve as image features for segmentation and registration, fiber strain responses as a result of mTBI are diffuse [20,34]. This makes it difficult to directly employ CNN-based techniques that are often built on local spatial filters (e.g., 3D CNN designed for tumor segmentation and measurement [35]).

Deep learning techniques are advancing rapidly. Instead of applying the most recent neural network models that are still under active development [31,35], here we chose a more conventional approach to first introduce this important research tool into the TBI biomechanics research field. Implicit features of the entire voxel-wise WM fiber strains were generated from reconstructed head impacts for concussion classification. Performances of the deep learning classifier were compared against baseline machine learning and univariate logistic regression methods in a leave-one-out cross-validation framework. This was important to ensure an objective comparison and to maximize rigor, which has often been overlooked in other biomechanical studies that only reported AUC from a single training dataset [8,12,13,18]. These injury prediction strategies are important extensions to previous efforts, which may provide important fresh insight into how best to objectively predict concussion in the future.

**Materials and Methods**
*The Worcester Head Injury Model (WHIM) and WM Fiber Strain*

We used the Worcester Head Injury Model (WHIM; **Fig. 1** [11,34]) to simulate the reconstructed NFL head impacts [36,37]. Descriptions of the WHIM development, material property and boundary condition assignment, and quantitative assessment of the mesh geometrical accuracy and model validation performances have been published previously. Briefly, the WHIM was created based on high resolution T1-weighted MRI of an individual athlete. DTI of the same individual provided averaged fiber orientations at each WM voxel location [11].

The 58 reconstructed head impacts include 25 concussions and 33 non-injury cases. Identical to previous studies [14,18,37,38], head impact linear and rotational accelerations were preprocessed before applying to the WHIM head center of gravity (CG) for brain response simulation. The skull and facial components were simplified as rigid-bodies as they did not influence brain responses.

Peak WM fiber strain, regardless of the time of occurrence during impact, was computed at each DTI WM voxel (N = 64272; [34]). For voxels not corresponding to WM, their values were padded with zeroes. This led to a full 3D image volume encoded with peak WM fiber strains (with surface rendering of the segmented WM shown in **Fig. 1c**). They served as classification features for deep neural network training and concussion prediction. The choice of fiber strain instead of more commonly used maximum principal strain was because of its potentially improved injury prediction performance [12,13,34]. As no neuroimages were available for the 58 impact cases, injury detection using a previous deep learning technique based on DTI parameters [33] was not applicable in this study.

*Deep Learning: Background*

Deep learning has dramatically improved the state-of-the-art in numerous research domains (see a recent review in Nature Methods [23]). However, its application in TBI biomechanics is nonexistent at present. This technique allows models composed of multiple processing layers to learn representations of data with multiple levels of abstraction [23]. A deep learning neural network uses a collection of logical units and their activation statuses to simulate brain function. It employs an efficient supervised update method [39] or an unsupervised network training strategy [40]. This makes it feasible to train a "deep" (e.g., more than 3 layers) neural network, which is ideal for learning large scale and high dimensional data.

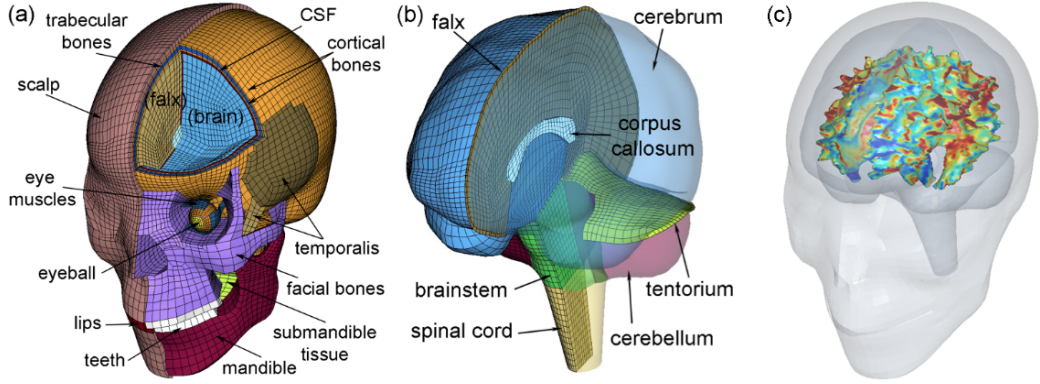

**Fig. 1** The WHIM head exterior (a) and intracranial components (b), along with peak fiber strain-encoded rendering of the segmented WM outer surface (c). The *x*-, *y*-, and *z*-axes of the model coordinate system correspond to the posterior–anterior, right–left, and inferior–superior direction, respectively. The strain image volume, which was used to generate the rendering within the co-registered head model for illustrative purposes, directly served as input signals for deep learning network training and concussion classification (see **Fig. 2**).

For a deep learning neural network, the *l*-th layer transforms an input vector from its lower layer, $a_{l-1}$, into an output vector, $a_l$, through the following forward transformation:

$$x_l = W_l a_{l-1} + b_l \tag{1}$$

$$a_l = \sigma_l(x_l) \tag{2}$$

where matrix $W_l$ is a linear transform describing the unit-to-unit connection between two adjacent, *l*-th and (*l*-1)-th, layers, and $b_l$ is a bias offset vector. Their dimensions are configured to produce the desired dimensionality of the input and output, with the raw input data represented by $x_0$ (**Fig. 2**). The nonlinear normalization or activation function, $\sigma_l$, can be defined as either a Sigmoid or a TanH function [41], or Rectified Linear Units (ReLU) [42] in order to suppress the output values for discriminant enhancement [43] and for achieving non-linear approximation [44]. Upon network training convergence, the optimized parameters, $W = \{W_l\}$ and $b = \{b_l\}$, are used to produce predictions of the cross-validation dataset. More details on the mathematics behind and procedures of deep network training are provided in the **Appendix**.

*Deep Learning: Network Design and Implementation*

A systematic approach to designing an "optimal" deep learning network is still an active research topic [45]. As a clear rule is currently lacking, trial-and-error is often used to determine the appropriate number of layers and the numbers of connecting units in each layer. Here, we empirically developed a network structure composed of five fully connected layers (i.e., each unit in a layer was connected to all units in its adjacent layers; **Fig. 2**), similarly to that used before [46]. The number of network layers was chosen to balance the trade-off between network structure nonlinearity and regularity.

The numbers of connecting units in each layer or the network dimension also followed a popular pyramid structure [46] to sequentially halve the number of connecting units in subsequent layers (i.e., a structure of 2000-1000-500-250 units for layers 1 to 4; **Fig. 2**). Each layer performed feature condensation transform (Eqns. 1 and 2) independently. The final feature vector, $x_5$, served as the input for injury classification. **Table 1** summarizes the dimensions of the weights, $W_l$, and offset vectors, $b_l$, as well as the normalization functions, $\sigma_l$, used to define the deep network. In total, the network contained over $1.31 \times 10^8$ independent parameters.

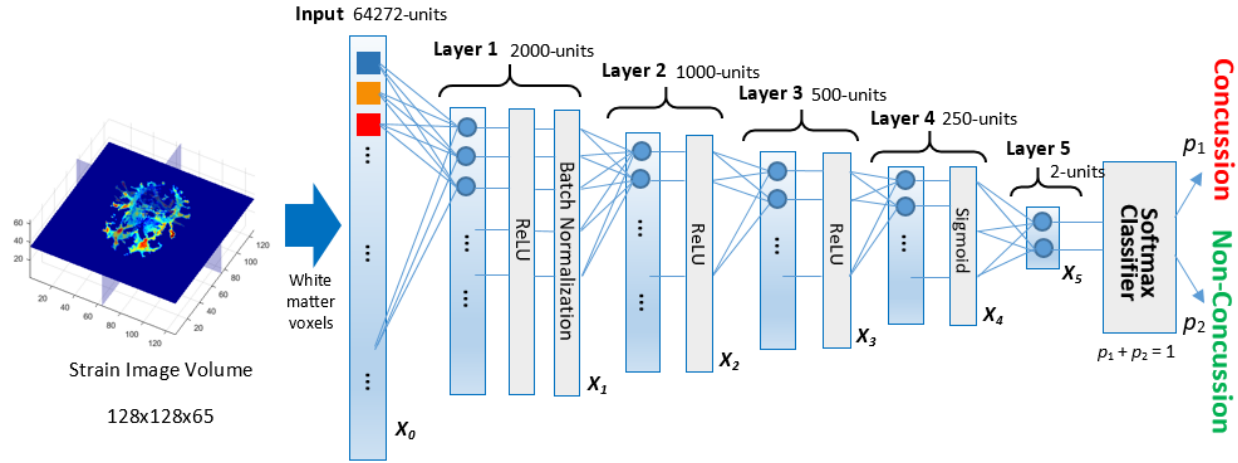

**Fig. 2** Structure of the deep learning network. The network contained five fully connected layers to progressively compress the fiber-strain-encoded image features, and ultimately, into a two-unit feature vector for concussion classification.

For the first three layers (i.e., layers 1 to 3 in **Fig. 2**), ReLU were used that provided a sparser activation than TanH and Signmoid functions to allow faster and more effective training [42]. A batch normalization technique was also used to avoid internal co-variate shift as a result of non-normal distributions of the input and output values. This enhanced the network robustness [47]. In contrast, the last layer prior to classification (layer 4 in **Fig. 2**) adopted a Sigmoid function to normalize output values to [0, 1], which was necessary to facilitate the Softmax classification [48].

**Table 1**. Summary of the dimensions of the weights and offset parameters, along with the normalization functions used to define the deep learning network. See Appendix for details regarding the normalization functions.

| Parameter | Definition |
|---|---|
| $W_l$: 2D matrix | $W_1$: 2000×64272; $W_2$: 1000×2000; $W_3$: 500×1000; $W_4$: 250×500; $W_5$: 20×250 |
| $b_l$: 1D vector | $b_1$: 2000 dim; $b_2$: 1000 dim; $b_3$: 500 dim; $b_4$: 250 dim; $b_5$: 2 dim |
| $\sigma_l$: normalization function | $\sigma_1$: ReLU + batch normalization; $\sigma_2$: ReLU; $\sigma_3$: ReLU; $\sigma_4$: Sigmoid; $\sigma_5$: no normalization (i.e., using an identity matrix) |

Upon training convergence, the initial high dimensional feature vector was condensed into a more compact representation. A Softmax function, $S$ (Eqn. A1 in **Appendix**), transformed the input feature vector, $y$, into a final output vector, $(p_1, p_2)$. The corresponding vector values represented the probability of concussion ($p_1$) and non-injury ($p_2$), respectively, where $p_1 + p_2 = 1$, by necessity. Concussion was said to occur when $p_1 \geq 0.5$.

The network was trained via an ADAM optimization [49] in Caffe [50]. A number of hyper-parameters needed to be optimized to achieve a satisfactory performance. With trial and error, we selected a gradient descent step size or learning rate of $2 \times 10^{-8}$ for all network layers, and the gradient descent momentum (i.e., the weight to multiply the gradient from the previous step in order to augment the gradient update in the current step) was set to 0.5. The default parameter values, $\beta_1 = 0.9$, $\beta_2 = 0.999$ and $\varepsilon = 10^{-8}$, were used to prevent the weights from growing too fast. The training dataset was divided into a batch size of 5 for training (randomly resampled cases were added when the remaining batch was fewer than 5). The maximum number of epochs was 5000 (an epoch is a complete pass of the full training dataset through the neural network).

*Early Stopping for Optimal Training*

An optimal number of training epochs achieves the best cross-validation accuracy at the minimum computational cost. However, this is not feasible to determine for fresh, unmet cases. Here, we monitored the validation accuracy of the training dataset to empirically determine a stopping criterion. Specifically, three training trials (in a leave-one-out cross-validation framework, see below) were generated to observe the convergence behaviors of the training and validation error functions (internally, 10% of the training dataset were used for validation within the deep learning training iterations; Eqn. A4 in **Appendix**; **Fig. 3**). The training error function asymptotically decreased with the increase in the number of epochs. The validation error function initially decreased, as expected, but started to increase after sufficient epochs, indicating overfitting has occurred.

These observations suggested the use of an "early stopping criterion" [48] to ensure sufficient training with a minimum number of epochs. Initially, 300 training epochs were empirically used to monitor the validation error convergence behavior [51]. If validation error did not decrease, the network training was considered as failed due to a poor initialization and the training would terminate. With the chosen learning rate $2\times10^{-8}$, we found that the network always converged within [1000, 5000] epochs, which was set as an admissible range of epochs. On the other hand, a larger learning rate often triggered overfitting. The empirical early stopping criterion allowed sufficient training while minimizing the risk of overfitting.

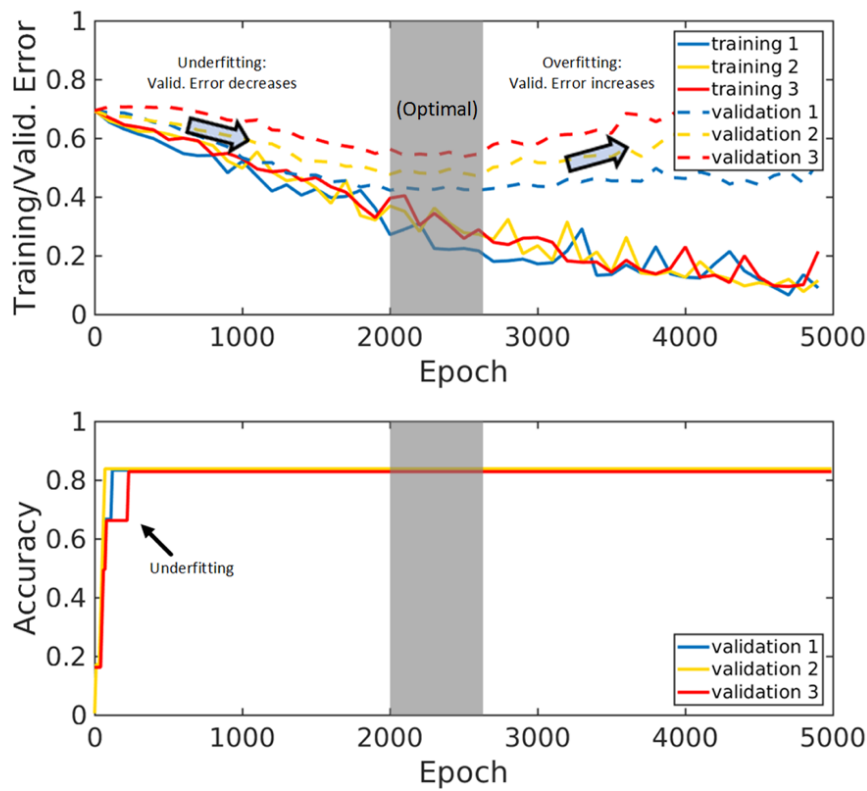

**Fig. 3** Illustration of training and validation error functions from three deep learning training trials (Top), along with the corresponding validation accuracy (based on the 10% training dataset used for validation internally; Bottom), vs. training epochs for three randomly generated trials. Maximum validation accuracies based on validation datasets were achieved using an early-stopping criterion after 2000 epochs.

*Concussion Classification and Performance Evaluation*

An objective evaluation of the concussion classification performances was important. Previously, a repeated random subsampling framework was employed to split the injury cases into independent and non-overlapping training and cross-validation datasets [14]. Given the relatively small sample size (N=58), here we adopted a leave-one-out cross-validation for performance evaluation. This maximized the training dataset so that to allow mimicking a real-world injury prediction scenario by potentially optimizing the prediction on a fresh, unmet head impact.

*Performance Comparison against Baseline Machine Learning Classifiers*

Conventional SVM and RF were selected as baseline classifiers to benchmark the performance of deep learning. Typically, a machine learning classifier requires an explicit feature selection to reduce input dimensionality and remove redundant, irrelevant, and noisy features from the input data in order to improve performance [52]. However, there is no standard approach for feature selection. For example, while the F-score approach is common for SVM [53], RF offers feature selection by itself [54]. In contrast, an explicit feature selection is not necessary in deep learning [55] as this is automatic during the optimization to maximize the input-output correlation. For completeness, here we conducted classification first without feature selection using the entire dataset as input to provide a reference performance for each classifier. After feature selection, they were compared in a more typical scenario for the two baseline machine learning techniques.

To avoid the classical feature selection bias problem [56], independent feature selections were performed for each leave-one-out cross-validation trial. Specifically, only the training dataset (N=57), not including the cross-validation data point, were used for feature selection, with either the F-score or RF-based approach. Using the recommended strategy [53], the F-score approach retained approximately 4% of features (N=2566; empirically determined to yield the highest cross-validation accuracy for SVM). For the RF-based method, a simplified variant of the conventional "gini" importance ranking approach [54] was used. A total of 5000 randomly initialized runs of RF were first conducted so that all of the voxels had a chance to serve as an important feature. After each run, the top 1% highest ranked features were retained to vote on a voxel-wise basis. The top 1% most frequently voted voxels (N=643) among all of the runs were finally selected. The "top 1% criteria" were similarly determined empirically to yield the highest cross-validation accuracy for RF. After the 58 independent feature selections, a probability map was generated based on the frequency of each WM voxel selected as an important feature for classification.

A linear kernel was used for SVM [53]. For RF, the numbers of decision trees and depths were determined empirically to maximize cross-validation accuracy. They were 45 and 64 without feature selection, and 75 and 8, or 75 and 12, respectively, when using the F-score or RF for feature selection. As RF depended on a random initialization, 100 RF trials were conducted for each training/injury prediction. For deep learning with feature selection, a smaller neural network with 5 fully connected layers of dimensions of 500-250-125-60-2 was designed to accommodate the substantially reduced feature size, which resulted in $4.85 \times 10^5$ independent parameters. The learning rate was adjusted to $1 \times 10^{-6}$. Other hyper-parameters remained unchanged.

*Performance Comparison against Scalar Injury Metrics*

In TBI biomechanics research, univariate logistic regression is the most commonly used method to report the AUC of a single training dataset [5,6,8,13,22]. They rely on a scalar response metric which is essentially a single, pre-defined feature. The following four injury metrics were used for further performance comparison: Brain Injury Criteria (BrIC [6]; a kinematic metric found to correlate the best with strain-based metrics in diverse automotive impacts [15]), CSDM for the whole brain (CSDM-WB) and the CC (CSDM-CC) based on maximum principal strain [57], as well as peak WM fiber strain in the corpus callosum (Peak-CC; [13,14]). The critical angular velocities for BrIC depend on the model used. For WHIM, they were 30.4 rad/s, 35.6 rad/s, and 23.5 rad/s along the three major axes, respectively [16]. For CSDM, an "optimal" strain threshold of 0.2 was used, which was to maximize the significance of injury risk-response relationship for the group of 50 deep WM regions using the same reconstructed NFL injury dataset [14].

Upon training convergence or after fitting, all classifiers generated a probability score for each of the impact case in the training and cross-validation datasets. For deep learning, this was $p_1$ in **Fig. 2** (Eqn. A1 in Appendix). This allowed constructing an ROC to report AUC (*perfcurve.m* in Matlab). For each classifier, an AUC for each training dataset was calculated based on 57 impact cases for each of the 58 independent injury predictions (as necessitated by the leave-one-out cross-validation framework). An average AUC was then reported. In contrast, a single AUC value for the testing dataset was obtained based on the probability scores of the 58 independent predictions.

**Data Analysis**

Simulating each head impact of 100 ms duration in Abaqus/Explicit (Version 2016; Dassault Systèmes, France) required ~50 min on a 12-CPU Linux cluster (Intel Xeon E5-2680v2, 2.80 GHz, 128 GB memory) with a temporal resolution of 1 ms. An additional 9 min was needed to obtain element-wise cumulative strains (single threaded). The classification framework was implemented on Windows (Xeon E5-2630 v3, 8 cores, 16 GB memory) with GPU acceleration (NVidia Titan X Pascal, 12 GB memory). Training each deep neural network typically required ~15 min and ~7 min for the two networks, respectively, but subsequent injury prediction was real-time (<0.01 sec).

For all the concussion classifiers, their performances were compared in terms of cross-validation accuracy, sensitivity, and specificity, as well as AUCs for both the training and testing datasets. All data analyses were conducted in MATLAB (R2016b; Mathworks, Natick, MA).

**Results**

*Strain-encoded whole-brain image volume*

**Fig. 4** illustrates and compares peak WM fiber-strain-encoded images on three orthogonal planes for a pair of striking and struck (non-injured and concussed, respectively) athletes involved in the same head collision. Without feature selection, deep learning directly utilized all of the strain-encoded WM image features for training and concussion classification.

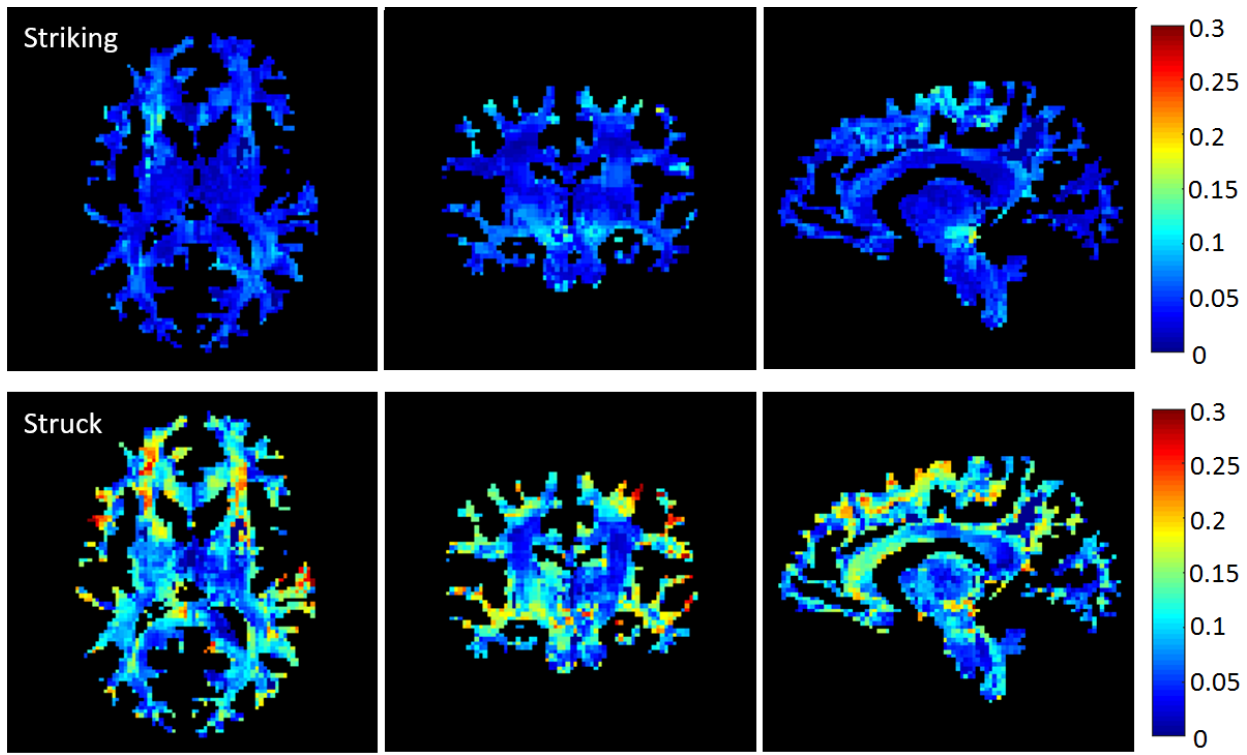

**Fig. 4** Cumulative WM fiber strains on representative orthogonal planes for a pair of striking (non-injury) and struck (concussed) athletes.

*Deep learning vs. SVM and RF*

Without an explicit feature selection, deep learning outperformed both SVM and RF in accuracy and specificity, with sensitivity slightly lower than that of RF (**Table 2**). SVM performed the worst in all categories. Feature selection improved the performances of both SVM and RF in all categories, regardless of the specific feature selection approach. However, only RF-based feature selection slightly improved the accuracy of deep learning, at the cost of slightly lowering specificity (**Table 3**). **Fig. 5** shows the probability maps indicating the frequency of each WM voxel serving as an important feature for classification using either the F-score or RF-based approach. Features identified by the former was more substantial because 4% of all WM voxels were selected from each trial, vs. only 1% for the latter method. For the RF-based method, the right superior longitudinal fasciculus (SLF-R) and left external capsule (EC_L) were two dominant regions often selected for classification.

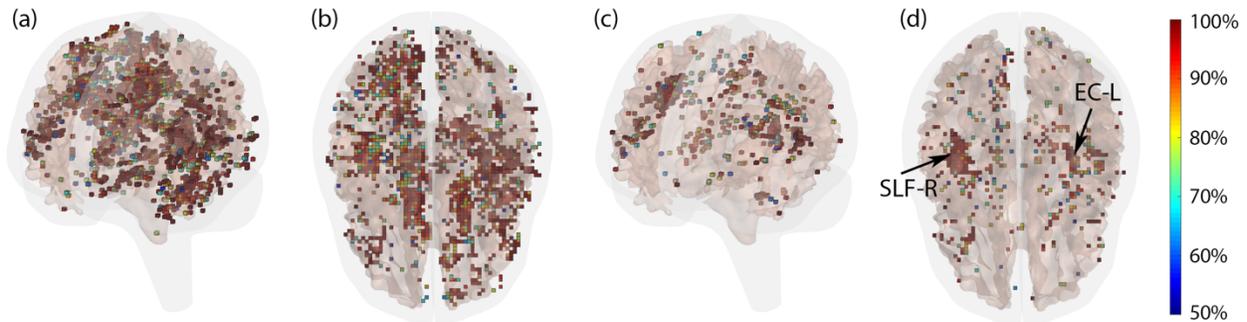

**Fig. 5**. Probability maps for WM voxels selected by the F-score (a and b) or RF-based (c and d) approach based on 58 independent feature selections. In each trial, the two approaches selected 4% and 1%, respectively, of the WM voxels as features. To improve visualization, only voxels with a probability greater than 50% (i.e., selected by at least 29 times) were shown. For the RF-based approach, SLF-R and EC-L were two dominant regions often selected for classification.

*Scalar injury metrics using univariate logistic regression*

The best performing deep learning, SVM and RF classifiers in terms of accuracy were obtained with RF-based feature selection (**Table 4**). They all had significantly higher performances in all categories than the scalar injury metrics from univariate logistic regression. Deep learning continued to perform the best among all classifiers in AUC using the testing dataset.

**Table. 2**. Summary of cross-validation accuracy, sensitivity, and specificity based on the testing dataset in a leave-one-out framework for deep learning, SVM and RF. No feature selection was conducted and WM voxels of the entire brain were used for classification. Results for RF were reported in the form of (mean±std) because 100 random trials were conducted for each prediction to accommodate the random initialization.

|  | **Deep learning** | **SVM** | **RF** |
|---|---|---|---|
| **Accuracy** | 0.845 | 0.724 | 0.811±0.023 |
| **Sensitivity** | 0.760 | 0.640 | 0.774±0.040 |
| **Specificity** | 0.909 | 0.788 | 0.839±0.030 |

**Table 3.** Performance summary of the three feature-based classifiers when using either the F-score or RF-based approach for feature selection prior to classification.

|  | Deep learning | | SVM | | RF | |
| --- | --- | --- | --- | --- | --- | --- |
|  | F-score | RF-feat | F-score | RF-feat | F-score | RF-feat |
| Accuracy | 0.845 | 0.862 | 0.828 | 0.828 | 0.828±0.018 | 0.842±0.016 |
| Sensitivity | 0.880 | 0.840 | 0.800 | 0.760 | 0.768±0.032 | 0.787±0.027 |
| Specificity | 0.818 | 0.879 | 0.848 | 0.879 | 0.873±0.019 | 0.883±0.021 |

**Table 4.** Performance summary of the best performing feature-based classifiers (all with RF-based feature selection) as well as of the four scalar metrics from univariate logistic regression. Accuracy, sensitivity, specificity and AUC were reported based on the 58 separate injury predictions in the leave-one-out cross-validation framework, along with the average AUC for the corresponding training datasets.

|  | Deep learning | SVM | RF | BrIC | CSDM-WB | CSDM-CC | Peak-CC |
| --- | --- | --- | --- | --- | --- | --- | --- |
| Accuracy | 0.862 | 0.828 | 0.842±0.016 | 0.776 | 0.741 | 0.776 | 0.690 |
| Sensitivity | 0.840 | 0.760 | 0.787±0.027 | 0.640 | 0.640 | 0.760 | 0.600 |
| Specificity | 0.879 | 0.879 | 0.883±0.021 | 0.879 | 0.818 | 0.788 | 0.758 |
| AUC-Testing | 0.892 | 0.872 | 0.856 | 0.781 | 0.786 | 0.771 | 0.737 |
| AUC-Training mean ± std (min, max) | 0.967±0.011 (0.930, 0.991) | 0.963±0.007 (0.948, 0.984) | 1.000±0.000 (1.000, 1.000) | 0.805±0.009 (0.797, 0.832) | 0.838±0.008 (0.831, 0.866) | 0.815±0.008 (0.807, 0.846) | 0.770±0.009 (0.760, 0.794) |

Finally, ROCs were produced for each classifier based on the testing dataset (**Fig. 6**). Two additional ROCs corresponding to the best and worst AUCs, respectively, were also produced for each classifier from the training datasets, as typically reported in other TBI biomechanical studies ([8,12,13,18]; **Fig. 7**).

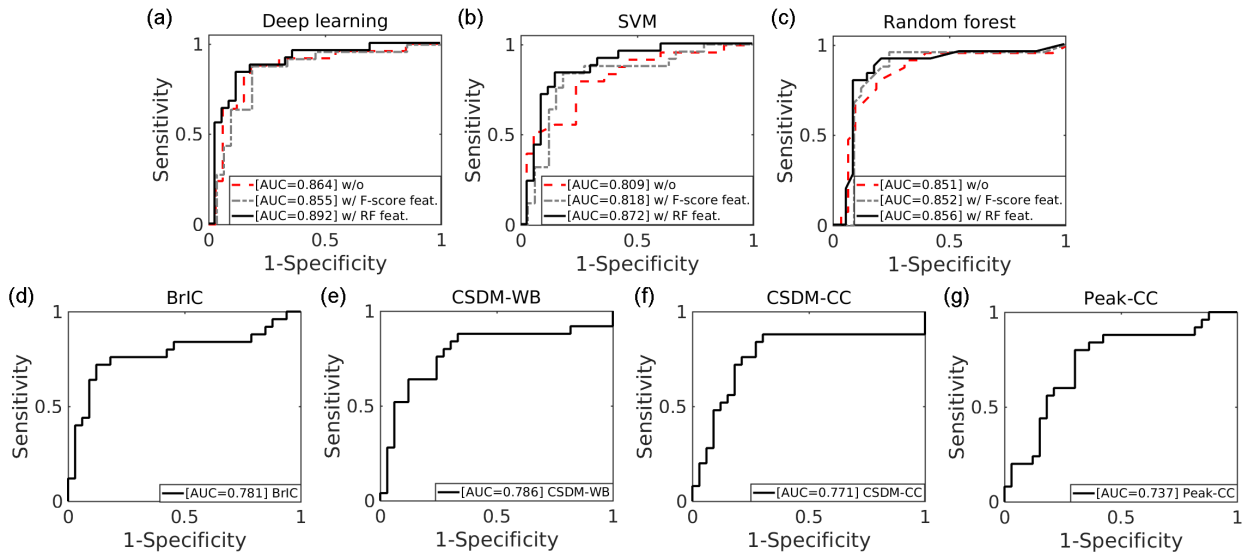

**Fig. 6** Comparisons of ROCs based on the testing dataset for the total of 7 classifiers.

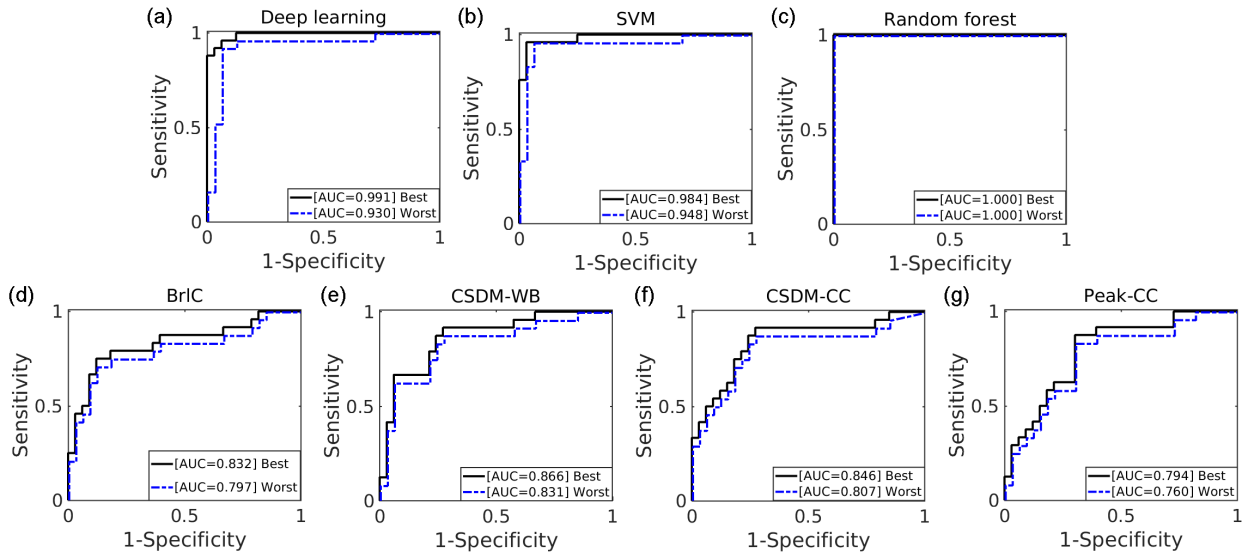

**Fig. 7** Similar comparisons of ROCs based on the training datasets. For the deep/machine learning techniques, only results from those with the RF-based feature selection are shown. The two ROCs correspond to the best and worst AUC, respectively.

**Discussion**

Developing an accurate and reliable injury predictor is one of the cornerstones in TBI biomechanics research for decades. Much of the work has so far focused on developing a single, scalar metric to describe impact severity and to predict injury. Numerous kinematics and model-estimated response variables have been proposed. Nonetheless, an "optimal" injury metric remains elusive and does not yet exist. However, a single scalar metric may not be sufficient for mTBI, including concussion, given the widespread neuroimaging alterations [20] and a diverse spectrum of clinical signs and symptoms [58] observed in the clinic. In this study, instead of similarly attempting to pinpoint an explicit response measure pre-defined in a specific ROI, we employed voxel-wise WM fiber strains from the entire brain as implicit features for injury prediction. The classical injury prediction was formulated into a supervised classification. Deep learning automatically distilled the most discriminative features from the strain-encoded image volumes for concussion classification. This was in sharp contrast to the current common approach in which a pre-defined scalar feature was essentially "hand-picked" to fit a univariate logistic regression model.

Based on 58 reconstructed NFL head impacts, we showed that the deep learning classifier significantly outperformed all of the four kinematic or response-based injury metrics selected here, in all of the performance categories (**Tables 2** and **4**). Only BrIC slightly outperformed deep learning in specificity when the F-score approach was used for feature selection (0.879 vs. 0.818; **Tables 3** and **4**). However, when no features were explicitly selected as in a more typical scenario, deep learning outperformed BrIC in all performance categories. The deep learning classifier also outperformed the two baseline machine learning classifiers in cross-validation accuracy regardless of whether features were first selected. Both the F-score and RF-based approaches improved the performances of SVM and RF. However, the latter was more effective for the RF classifier with increased accuracy, sensitivity and specificity (**Table 3**). The RF-based feature selection also improved the accuracy and sensitivity for the deep learning classifier, but at a cost of lowering specificity (**Tables 2** and **3**).

With the RF-based feature selection, all of the feature-based classifiers outperformed the scalar metrics using univariate logistic regression, for all of the performance categories (**Table 4**). In terms of AUC, which is widely used in TBI biomechanics research [8,12,13,18], the training dataset consistently generated larger scores than their counterparts using the testing dataset (**Table 4**), with RF even achieving a perfect AUC score of 1.0 (**Fig. 7**). All feature-based classifiers significantly outperformed the scalar metrics using either the testing (**Fig. 6**) or training (**Fig. 7**) dataset (average AUC of 0.873 vs. 0.769 for the testing dataset, vs. AUC of 0.977 and 0.807 for the training dataset, respectively). However, deep learning achieved the highest AUC based on testing dataset (**Table 4**; **Fig. 6**).

The highest AUC from a single training dataset using the latest KTH model (of 0.9655, when using peak WM fiber strain in the brainstem serving as the predictor [13]) was comparable to that of the three feature-based predictors reported here (range of 0.963–1.000; **Table 4**). However, no objective performance comparison can be made here as no cross-validation was performed in that study.

*Feature selection*

Deep learning does not typically require an explicit feature selection [55], as it is performed implicitly during the iterative training. However, the RF-based feature-selection approach did improve the deep learning classifier accuracy and sensitivity (albeit, at a cost of slightly lower specificity). Likely, this was an indication that the adopted deep learning classifier architecture (**Fig. 2**) may not be optimal. It was possible to further improve the deep neural network architecture and fine-tune the hyper-parameters. However, this may be ill-advised. First, a clear guideline is currently lacking on how best to design the deep neural network architecture. Therefore, an exhaustive trial-and-error effort would be necessary to achieve the absolute best performance, and increasing the number of neural network layers or layer-wise units would lead to challenges in computational cost and memory requirement. More importantly, it is known that the reconstructed injury dataset may have errors in impact kinematics [36] and it suffers from the under-sampling of non-injury cases [14]. Therefore, an "optimal" neural network model with the "best" performance may not be applicable when it is applied to a more typical general population. The need to further cross-validate an injury predictor using a separate dataset was recently explored [59]. This will be a topic of further research in the future.

Feature selection was important for both SVM and RF, without which SVM had a rather poor performance. This was likely a typical "curse of dimensionality" due to the small sample size that led to data overfitting [53], especially since a simple linear kernel was used for classification. Both feature selection methods were effective in improving performance. The RF-based approach consistently identified regions in SLF-R and EC-L as important classification features (**Fig. 5**). Incidentally, SLF-R was also found to be one of the most injury discriminative ROIs based on injury susceptibility measures via logistic regression [14]. The consistency here suggested concordance between the different classification approaches based on the same dataset. However, caution must be exercised when attempting to extrapolate this finding to other subject groups, particularly given that neuroimages corresponding to a single subject were used here for the group of subjects that did not account for individual variability. A subject-specific study would be desirable to address these limitations in the future, which was not feasible here.

*Feature-based classifiers vs. scalar injury metrics*

Feature-based machine/deep learning classifiers utilized multiple features for classification. It started from the entire voxel-wise WM fiber strains. With data-driven feature-selection aimed at reducing redundant information in the input and to avoid data overfitting, multiple features were retained for subsequent classification to maximize performance. In contrast, scalar injury metrics relied on a single response variable often empirically pre-defined. Kinematic injury metrics, including BrIC [6], are typically constructed by using the peak magnitudes of linear/rotational acceleration or velocity, and their variants. They characterize impact severity to the whole brain, and are unable to provide tissue response directly. While a head FE model estimates tissue responses throughout the brain, only the peak response magnitude of a single element in a pre-defined ROI (e.g., peak-CC [13]) or a dichotomous volume fraction above a certain threshold (e.g., CSDM-WB [57] and CSDM-CC) is used for injury prediction. Similar to kinematic injury metrics, critical information is lost on the location or distribution of peak brain responses, even though such information is already available. Because of these inherent limitations with scalar injury metrics, it was not surprising that all of the feature-based classifiers significantly outperformed all of the scalar injury metrics, regardless of the performance category (e.g., average accuracy of 0.844 vs. 0.746, and average AUC from the testing dataset of 0.873 vs. 0.769, respectively; **Table 4**).

Compared with scalar injury metrics, deep learning was the extreme opposite as it utilized information from all of the WM voxels of the entire brain as input for classification. The technique has also been successfully applied to three-dimensional neuroimages for injury and severity detection [33]. Conceivably, this may enable a multi-modal injury prediction combining both biomechanical responses (e.g., strain-encoded image volume in **Fig. 4**) and corresponding neuroimages such as DTI of the same subjects to improve injury prediction performance. This is beyond the capabilities of any kinematic or strain-based injury metrics currently in use.

In addition, a strain threshold was necessary to dichotomize the brain ROI volumes for CSDM measures. An "optimal" strain threshold of 0.2 was previously determined by maximizing the significance of risk-response relationship for the group of 50 deep WM ROIs [14]. While adjusting the strain threshold could provide additional fitting flexibility to further improve the injury prediction performances of the scalar injury metrics, it may also lead to inconsistencies in threshold when each individual ROIs were used for injury prediction. Similarly to the ill-advised effort in reaching the absolute "best" performance with deep learning, this is undesirable, as the strain threshold is related to the physical injury tolerance found from actual *in vivo/in vitro* injury experiments. Importantly, deep learning and the two baseline machine learning classifiers have consistently outperformed all scalar injury metrics using univariate logistic regression. Therefore, this suggests strong motivation for further investigation into the use of the more advanced feature-based concussion classifiers in the future.

*Comparison with previous findings*

With the same injury dataset, Zhao and co-workers analyzed the injury susceptibilities and vulnerabilities of the entire deep WM ROIs and neural tracts [14]. A univariate logistic regression of each individual ROI/neural tract was conducted to report accuracy, sensitivity, specificity, and training AUC averaged from 100 trials in a repeated random subsampling cross-validation framework. A direct comparison was not feasible here because a leave-one-out cross-validation scheme was adopted in this study instead. Nevertheless, deep learning continued to outperform or at least to be comparable to the performances of each individual ROI/neural tract (e.g., accuracy of 0.862 with deep learning vs. 0.852 using point-wise injury susceptibility in SLF-R). However, unlike the previous study that required registering the FE model to a WM atlas to identify ROIs/neural tracts, no registration or segmentation was necessary with deep/machine learning that used the entire WM voxels as input. In addition, the previous study relied on dichotomized injury susceptibilities, which depended on a strain threshold similarly to the CSDM metrics selected here. This was unnecessary with deep/machine learning.

Another study identified Peak-CC to considerably outperform BrIC in AUC using all of the reconstructed NFL impacts as a single training dataset (0.9488 vs. 0.8629 [13]). Here we reported the opposite (average AUC of 0.770 vs. 0.805 for Peak-CC and BrIC in the training dataset, respectively; **Table 4**). This suggested disparities between the two head injury models and their analysis approaches. Perhaps most notably, the two models differ in material properties (isotropic, homogeneous vs. anisotropic for the WM). In addition, they have different brain-skull boundary conditions (nodal sharing via a soft layer CSF vs. frictional sliding), mesh resolution (average size of 3.2 mm vs. 5.8 mm), method to calculate fiber strain (projection of a strain tensor vs. assigning averaged fiber directions directly to FE elements), and even segmentation of the CC [11].

Nevertheless, improving a model's injury predictive power is a constant process. Together with more well-documented real-world injury cases, further comparison of injury prediction performances across models is important to understand how best to improve. A high AUC in a training dataset does not necessarily indicate the same high level of AUC or other performance categories using the testing dataset. For example, CSDM-WB had a higher AUC in training (average value of 0.838, vs. 0.805 and 0.815 for BrIC and CSDM-CC; **Table 4**), but it performed worse in cross-validation accuracy (0.714 with CSDM-WB vs. 0.776 for BrIC and CSDM-CC). Therefore, it is important that future studies utilize cross-validation, rather than training or fitting, performances for objective evaluation and comparison.

*Limitations*

The superior performances of the deep learning and baseline machine learning classifiers were encouraging. However, it must be recognized that only one head FE model and a single injury dataset were employed here for performance evaluation and comparison. As even validated head models could produce discordant brain responses [60], it is important to further evaluate whether similar performance gains are possible with estimated brain responses from other head injury models. In addition, errors in the reconstructed head impact kinematics [36] are well-known, and the resulting uncertainties in model results, and implications in injury prediction due to under-sampling of non-injury cases [13,18,22] have been extensively discussed. Further, this dataset does not consider the cumulative effects from repetitive sub-concussive head impacts, the importance of which is becoming realized. Therefore, the deep learning classifier trained here may not be readily applicable to other injury datasets and a fresh training is necessary.

Importantly, a feature-based deep learning classifier has not been applied to TBI biomechanics before, despite its numerous recent successes across a wide array of scientific domains [23]. The deep learning approach and cross-validation framework established here may set the stage for continual development and optimization of a response-based injury predictor in the future. With further cross-validation using more independent injury datasets, the value of deep learning in TBI biomechanical investigations will be better studied.

Nevertheless, limitations with deep learning are also noted. First, empirical experience is often necessary to design the network structure, as a clear guideline is lacking. The fact that RF with feature selection outperformed deep learning in sensitivity (when no features were explicitly selected) may indicate that the deep neural network architecture may not be optimal, and there could still be room for improvement. In addition, unlike scalar injury metrics relying on explicit features, deep learning behaves much like a "black box" without an obvious physical interpretation of the its internal decision mechanism. Therefore, although an explicit feature selection was not necessary with deep learning, it may still be valuable to provide insight into the most injury discriminative features. In addition, the resulting reduction in feature size would also improve computational efficiency (15 min vs. 7 min for training).

Finally, the limitation of WHIM using isotropic, homogeneous material properties of the brain was discussed [34]. In addition, a generic head model and the corresponding neuroimages of one individual, rather than subject-specific head models and individualized neuroimages, were used to study a group of athletes. Inter-subject variation in neuroimaging and uncertainty in strain responses on an individual basis could not be evaluated. Nevertheless, a generic model is a critical steppingstone towards developing individualized models and to couple with their own neuroimages for more personalized investigations in the future. This is analogous to the typical $50^{th}$ percentile head models currently in use that do not yet directly correspond to detailed neuroimages [14].

**Conclusion**

We introduced a deep learning classifier into biomechanical investigations of traumatic brain injury. The technique utilized voxel-wise white matter fiber strains of the entire brain as input for concussion prediction. Based on reconstructed NFL head impacts, we showed that feature-based classifiers, including deep learning and two baseline machine learning classifier, outperformed all of the four selected scalar injury metrics in all performance categories in a leave-one-out cross-validation framework. Deep learning also achieved higher performances than the two baseline machine learning techniques in cross-validation accuracy, sensitivity, and AUC. The deep neural network developed here was by no means optimal or was ready for deployment in a more typical, general population. Nevertheless, the superior performances of deep learning and conventional feature-based machine learning in concussion prediction, especially relative to the commonly used scalar injury metrics via univariate logistic regression, suggest its promise for future applications in biomechanical investigations of traumatic brain injury.

**Acknowledgements**

Funding is provided by the NIH grants R01 NS092853 and R21 NS088781. The authors are grateful to the National Football League (NFL) Committee on Mild Traumatic Brain Injury (MTBI) and Biokinetics and Associates Ltd. for providing the reconstructed head impact kinematics. The Titan X Pascal used for this research was donated by the NVIDIA Corporation.

**Compliance with ethical standards**

**Conflict of interest:** We have no competing interests.

**Appendix: Deep Learning Backpropagation for Supervised Training**

An objective error function from the previous network layer (see **Fig. 2**) can be used to maximize the input-output correlation either in an unsupervised [61] or a supervised [62] manner to minimize training error. Here, we used a supervised method for concussion classification, as supported by Caffe [50]. A Softmax classifier [48] based on condensed feature vector was adopted. Mathematically, this classifier is defined as:

$$S_x(j) = \frac{e^{x(j)}}{\sum_k e^{x(k)}} \tag{A1}$$

where $x(j)$ and $x(k)$ are the $j$-th and $k$-th element of the feature vector, $x$, respectively, obtained from the trained network (output from the final layer). The classifier was trained by minimizing the Cross-Entropy error function relative to the known data label, $t(k)$, of either 0 or 1 (representing concussion or non-injury, respectively, in our study) for a training dataset, $x$, and its corresponding classifier output, $S_x$ [63]:

$$E(x) = -\sum_k [t(k) \log S_x(k) + (1 - t(k)) \log(1 - S_x(k))] \tag{A2}$$

The total error, $E = \sum_x E(x)$, for the training dataset served as the objective function for training via a backpropagation algorithm, as described below.

For a deep learning network with parameters, $W = \{W_l\}$ and $b = \{b_l\}$, the error function in Eqn. A2 can be represented as $E(W, b)$, which quantifies the classification error between the predicted and ground-truth labels. Deep network training is to optimize $W$ and $b$ in order to minimize the error, $E$. An efficient approach is through a backpropagation algorithm [62]. First, the network performs a *forward propagation* (Eqns. 1 and 2) to produce classification and the error function value. For a network of $L$ layers, the gradient of the error function with respect to $x_l$ at the $l$-th layer ($l \leq L$), $\delta_l = \nabla_{x_l} E$, can be iteratively computed via the following *backpropagation*:

$$\delta_L = \nabla_{a_L} E \odot \sigma'_L(x_L) \tag{A3}$$

$$\delta_l = W_{l+1}^T \delta_{l+1} \odot \sigma'_l(x_l) \tag{A4}$$

where $\odot$ is the element-wise product. These gradients are used to minimize $E$ via a gradient descent algorithm. Eqn. A3 and A4 are derived by the chain rule in calculus, and the mathematical details can be found in standard neural network textbook (e.g., [48] Chap. 4.7). After computing $\delta_l$, the gradients with respect to $W$ and $b$ are finally obtained:

$$\nabla_{b_l} E = \delta_l \tag{A5}$$

$$\nabla_{W_l} E = \delta_l \otimes a_{l-1}^T \tag{A6}$$

where $\otimes$ represents the tensor product. The following pseudo algorithm describes the training process for a network of $L$ layers.

1. Input training set $X$;
2. For each training sample $x$ in $X$:
    a. Compute the forward transformations (Eqns. 1 and 2) from layer 2 to $L$
    b. Compute the error, $\delta_L(x)$, (Eqn. A3) for layer $L$
    c. Compute backpropagation, $\delta_l(x)$, (Eqn. A4) from layer $L$ to 2
3. Gradient descent (for a given step size, $\lambda > 0$):
    a. Update $W$: $W_l \leftarrow W_l - \frac{\lambda}{|X|} \sum_x \delta_l(x) \otimes a_{l-1}^T(x)$ from Layer $L$ to 2
    b. Update $b$: $b_l \leftarrow b_l - \frac{\lambda}{|X|} \sum_x \delta_l(x)$ from Layer $L$ to 2

The training continues until the network is converged to generate optimized network parameters, $W$ and $b$, which are then fixed to perform classification on the cross-validation dataset.